\documentclass[11pt]{article}

\usepackage[utf8]{inputenc}
\usepackage[margin=1in]{geometry}
\usepackage{amsmath,amssymb}
\usepackage{graphicx}
\usepackage{booktabs}
\usepackage{caption}
\usepackage[affil-it]{authblk}
\usepackage{url}
\usepackage[colorlinks=true,linkcolor=blue,citecolor=blue,urlcolor=blue]{hyperref}

\newcommand{\Pmiss}{P_{\mathrm{miss}}}
\newcommand{\prev}{p_{\mathrm{review}}}
\newcommand{\tAI}{\tau_{\mathrm{AI}}}
\newcommand{\cFP}{c_{\mathrm{FP}}}
\newcommand{\cFN}{c_{\mathrm{FN}}}

\title{\textbf{Toward Responsible AI-Augmented Cyber Defense:\\
Pattern Recognition, Defense-in-Depth, and the Case for Human--AI Collaboration}}

\author[1]{Mustafa S. Aljumaily}
\author[2]{Hayder Kareem Abed}
\author[3]{Nawar S. Alseelawi}
\affil[1]{Research and Development (R\&D) Department, Daw Alfada Company, Baghdad, Iraq\\ \texttt{mustafa.s@daw-alfada.com}}
\affil[2]{CEO, Daw Alfada Company, Baghdad, Iraq\\ \texttt{haider.k@daw-alfada.com}}
\affil[3]{University of Misan, Maysan, Iraq\\ \texttt{nawar.alseelawi@uomisan.edu.iq}}
\date{}

\begin{document}

\maketitle

\begin{abstract}
\noindent
Cybersecurity literature has extensively documented the operational benefits of artificial intelligence (AI) for threat detection, incident response, and prevention, while raising qualitative concerns about over-automation, algorithmic bias, and analyst-skill erosion. What remains largely absent is a formal, falsifiable model connecting three constructs that recur across this literature: Defense-in-Depth Theory, the Artificial Intelligence Theory of Pattern Recognition, and human--AI collaboration in security operations. This paper develops such a model. We formalize layered defense as a Bernoulli detection cascade in which AI augmentation enters multiplicatively across layers; we formalize each layer's pattern-recognition behavior as a Neyman--Pearson/Bayesian detector with a derived closed-form optimal threshold; and we formalize human--AI triage as a capacity-constrained cascade with an explicit, quantifiable trade-off between detection probability and false-alarm (``alert fatigue'') rate. A Monte Carlo/analytical simulation evaluated at illustrative but realistic operating points shows that (i) AI augmentation compounds across defense layers, delivering its largest marginal gains exactly where traditional layering saturates, and (ii) full human review of AI-flagged alerts is not optimal: increasing analyst capacity toward 100\% coverage cuts false alarms by roughly 20-fold but simultaneously lowers system-level detection probability, because imperfect analyst accuracy is then applied to every alert rather than a filtered subset. These results give the widely repeated qualitative recommendation of ``balanced human--AI collaboration'' a precise, testable form and suggest an interior-optimum capacity ratio as a concrete design target for security operations centers (SOCs), including those securing IT/OT-converged critical infrastructure.

\medskip
\noindent\textbf{Keywords:} Artificial Intelligence; Cybersecurity; Defense-in-Depth; Pattern Recognition; Human--AI Collaboration; Bayesian Detection; Security Operations Center; Alert Fatigue.
\end{abstract}

\section{Introduction}\label{sec:intro}

The cybersecurity threat landscape has grown in scale and technical sophistication faster than traditional, rule-based defenses can adapt, a pattern documented across recent reviews of AI-driven security \cite{ejeofobiri2024,ejjami2024}. Two theoretical constructs recur throughout this literature as organizing frameworks: Defense-in-Depth Theory, which holds that layering independent security controls narrows the attack surface more than any single control could; and the Artificial Intelligence Theory of Pattern Recognition, which holds that machine learning systems can detect deviations from learned normal behavior that static, signature-based systems cannot \cite{ejjami2024}. Both theories are used descriptively---to organize a narrative account of where AI helps---but neither has, to our knowledge, been operationalized as a quantitative model that predicts how much a given layering strategy, AI augmentation level, or human-review policy will change system-level security outcomes.

This gap matters because the same literature identifies a genuine, unresolved tension: AI improves detection speed and coverage, but over-reliance on automation is repeatedly flagged as a risk to analyst skill retention and to the diversity of judgment a security operations center (SOC) can bring to novel threats \cite{ejeofobiri2024,ejjami2024}. Recommendations to ``balance automation with human oversight'' appear throughout, but without a model that specifies what balance means quantitatively, such advice is difficult to act on or to falsify.

This paper addresses that gap with three contributions:

\begin{enumerate}
\item A formal Defense-in-Depth cascade model in which AI-driven pattern recognition augments each layer's detection probability multiplicatively across the miss-probability product, giving a precise account of why AI's marginal value is largest where layering alone saturates (Sections~\ref{sec:cascade}--\ref{sec:augment}).
\item A Neyman--Pearson/Bayesian formalization of the pattern-recognition layer itself, including a closed-form cost-weighted optimal threshold, connecting the qualitative ``AI detects anomalies'' claim to standard detection theory (Section~\ref{sec:np}).
\item A capacity-constrained human--AI triage model that produces a specific, counterintuitive, and testable prediction---that full human review of AI-flagged alerts is not detection-optimal---replacing the general ``human oversight is good'' recommendation with an interior-optimum design target (Section~\ref{sec:triage}, Section~\ref{sec:results-triage}).
\end{enumerate}

The model is validated with an analytical/Monte Carlo simulation using illustrative Gaussian-separated detectors representative of a network intrusion detection system (IDS), an endpoint detection and response (EDR) layer, and an OT/SCADA anomaly monitor---a layer combination directly relevant to IT/OT-converged critical infrastructure such as oil and gas SCADA environments.

\section{Related Work}\label{sec:related}

Recent integrative and systematic reviews converge on three application areas for AI in cybersecurity \cite{ejeofobiri2024,ejjami2024,sarker2021}. First, AI-driven threat detection extends signature-based intrusion detection systems (IDS) with supervised and unsupervised machine learning---random forests, support vector machines, and neural networks for classification; clustering and principal component analysis for outlier detection---and deep learning architectures (CNNs, RNNs/LSTMs) for both network-traffic and malware analysis \cite{ejeofobiri2024,ahmad2021,liu2019}. Second, AI-powered response systems automate parts of the incident-response lifecycle through Security Orchestration, Automation, and Response (SOAR) platforms and AI-driven threat hunting, reducing the time between detection and containment \cite{ejeofobiri2024}. Third, AI-enhanced prevention techniques---predictive threat intelligence, AI-assisted vulnerability management, and behavioral biometrics---shift security posture from reactive to proactive \cite{ejeofobiri2024,ejjami2024}.

Running alongside these operational gains, both reviews raise a consistent set of concerns: adversarial manipulation of AI models, data-quality and representativeness problems that produce biased detection outcomes, the opacity (``black-box'' nature) of AI decision processes, and the risk that operational reliance on AI erodes human analytical skill over time \cite{ejeofobiri2024,ejjami2024}. Ejjami \cite{ejjami2024} frames these concerns within two theories---Cybersecurity Defense-in-Depth Theory and the Artificial Intelligence Theory of Pattern Recognition---and argues that responsible AI deployment requires combining AI's predictive capability with human oversight, transparency (explainable AI), and equitable access, particularly for smaller organizations with constrained AI budgets.

These reviews are, by design, qualitative syntheses: they identify recurring themes and theoretical constructs across a large literature rather than deriving quantitative predictions from them. Neither offers a mechanism for computing how much detection improvement a given layering-and-AI-augmentation strategy should be expected to deliver, nor a decision rule for setting a human-review capacity that is provably better than full automation or full human review. This paper takes the two theoretical constructs those reviews foreground and gives each a formal, testable mathematical structure, which we then validate computationally.

\section{Theoretical Framework}\label{sec:framework}

We adopt the same two guiding theories used in the reviewed literature and make each one operational:

\subsection{Cybersecurity Defense-in-Depth Theory}\label{sec:did-theory}

Defense-in-Depth Theory holds that layering independent, heterogeneous security controls (network, endpoint, application, and---in industrial environments---OT/SCADA-specific controls) reduces overall risk more effectively than optimizing any single control, because an attacker must evade every layer to succeed \cite{ejjami2024,nsa_did}. We formalize this as a probabilistic miss-cascade in Section~\ref{sec:cascade}.

\subsection{Artificial Intelligence Theory of Pattern Recognition}\label{sec:pr-theory}

This theory holds that AI systems can learn the statistical structure of ``normal'' behavior from data and flag deviations from it, enabling detection of threats that do not match any known signature \cite{ejjami2024}. We formalize this as a binary hypothesis test on a learned anomaly score, following standard Neyman--Pearson/Bayesian detection theory, in Section~\ref{sec:np}.

\subsection{Synthesis: Human--AI Collaboration as a Triage Cascade}\label{sec:synthesis}

Neither theory, individually, addresses where human judgment enters the pipeline or how its capacity should be allocated. We treat human--AI collaboration as a second, downstream cascade stage---an AI-prioritized triage queue with a capacity-constrained human reviewer---and derive its performance characteristics in Section~\ref{sec:triage}. This synthesis is the paper's central theoretical contribution: it connects Defense-in-Depth and Pattern Recognition Theory to a third, previously under-formalized construct using the same probabilistic language.

\section{Mathematical Model}\label{sec:model}

\subsection{Notation}\label{sec:notation}

Let $L$ be the number of independent defense layers indexed $i \in \{1,\dots,L\}$ (e.g., network IDS, endpoint EDR, OT/SCADA anomaly monitor). Let $H_0$ and $H_1$ denote the benign and attack hypotheses, with prior $\pi_1 = P(H_1)$. Each layer produces an anomaly score $s_i$ and applies a decision threshold $\tau_i$. $d_i(\tau_i) = P(s_i > \tau_i \mid H_1)$ is the layer's true-positive (detection) rate and $f_i(\tau_i) = P(s_i > \tau_i \mid H_0)$ is its false-positive rate. $\alpha_i \in [0,1]$ is an AI augmentation factor at layer $i$. $C$ is the human analyst's review capacity (alerts per unit time), and $d_H$, $f_H$ are the analyst's own detection/false-positive rates on reviewed alerts.

\subsection{Defense-in-Depth as a Layered Bernoulli Cascade}\label{sec:cascade}

Treating the system as detecting an attack whenever any layer flags it, and assuming layer-conditional independence given $H_1$:
\begin{equation}\label{eq:miss}
P(\text{system misses attack}) \;=\; \prod_{i=1}^{L} \bigl(1 - d_i(\tau_i)\bigr)
\end{equation}
\begin{equation}\label{eq:pd}
P_D(L) \;=\; 1 - \prod_{i=1}^{L} \bigl(1 - d_i(\tau_i)\bigr)
\end{equation}
\begin{equation}\label{eq:pf}
P_F(L) \;=\; 1 - \prod_{i=1}^{L} \bigl(1 - f_i(\tau_i)\bigr)
\end{equation}

$P_D(L)$ increases monotonically in $L$ with diminishing marginal returns, since each new layer only needs to catch what prior layers missed. $P_F(L)$, however, also increases with $L$---the formal counterpart of the ``alert fatigue'' problem noted qualitatively in the reviewed literature \cite{ejeofobiri2024,ejjami2024} and documented empirically in operational SOC settings \cite{hassan2019nodoze}---motivating the triage stage introduced in Section~\ref{sec:triage}. Structurally, Equations~\eqref{eq:miss}--\eqref{eq:pf} instantiate a classical distributed-detection architecture with an OR fusion rule \cite{varshney1997}; the contribution here lies in coupling that structure to AI augmentation (Section~\ref{sec:augment}) and to capacity-constrained human triage (Section~\ref{sec:triage}).

\subsection{Pattern Recognition as a Neyman--Pearson / Bayesian Detector}\label{sec:np}

Each layer's anomaly score is modeled as $s_i \mid H_0 \sim \mathcal{N}(\mu_{0i}, \sigma_{0i}^2)$ and $s_i \mid H_1 \sim \mathcal{N}(\mu_{1i}, \sigma_{1i}^2)$, with $\mu_{1i} > \mu_{0i}$. For a threshold rule (flag if $s_i > \tau_i$):
\begin{equation}\label{eq:rates}
d_i(\tau_i) \;=\; 1 - \Phi\!\left(\frac{\tau_i - \mu_{1i}}{\sigma_{1i}}\right),
\qquad
f_i(\tau_i) \;=\; 1 - \Phi\!\left(\frac{\tau_i - \mu_{0i}}{\sigma_{0i}}\right)
\end{equation}
where $\Phi$ is the standard normal CDF. Sweeping $\tau_i$ traces the layer's ROC curve; its area (AUC) summarizes discriminative quality \cite{fawcett2006}. By the Neyman--Pearson lemma \cite{neyman1933}, the cost-optimal decision rule is a likelihood-ratio test; for equal-variance Gaussians ($\sigma_{0i} = \sigma_{1i} = \sigma_i$) this reduces to a closed-form threshold:
\begin{equation}\label{eq:threshold}
\tau_i^{*} \;=\; \frac{\mu_{0i} + \mu_{1i}}{2} \;+\; \frac{\sigma_i^{2}}{\mu_{1i} - \mu_{0i}}\,\ln\eta,
\qquad
\eta \;=\; \frac{\cFP\,\pi_0}{\cFN\,\pi_1}
\end{equation}
where $\cFP$ and $\cFN$ are the relative costs of a false alarm versus a missed attack. Equation~\eqref{eq:threshold} gives a principled, cost-aware way to set each layer's operating point, directly addressing the reviewed literature's concern that skewed or arbitrary thresholds produce disproportionate false positives or negatives \cite{ejjami2024}.

\subsection{AI Augmentation of Layer Detection}\label{sec:augment}

Let $d_i^{0}$ denote a layer's baseline (signature/rule-based) detection rate. AI-driven pattern recognition recovers a fraction $\alpha_i$ of the attacks the baseline layer would otherwise miss:
\begin{equation}\label{eq:aug}
d_i^{\mathrm{AI}} \;=\; d_i^{0} + \bigl(1 - d_i^{0}\bigr)\alpha_i
\;\;\Longleftrightarrow\;\;
1 - d_i^{\mathrm{AI}} \;=\; \bigl(1 - d_i^{0}\bigr)\bigl(1 - \alpha_i\bigr)
\end{equation}
Substituting into Equation~\eqref{eq:miss}:
\begin{equation}\label{eq:compound}
\Pmiss^{\mathrm{AI}}(L) \;=\; \prod_{i=1}^{L} \bigl(1 - d_i^{0}\bigr)\bigl(1 - \alpha_i\bigr)
\;=\; \Pmiss^{0}(L) \cdot \prod_{i=1}^{L}\bigl(1 - \alpha_i\bigr)
\end{equation}
Equation~\eqref{eq:compound} is the paper's central structural claim about AI-augmented Defense-in-Depth: AI's contribution compounds multiplicatively across layers, so even modest per-layer gains ($\alpha_i \approx 0.3$--$0.4$) yield large system-level improvements once $L \geq 2$--$3$. This gives quantitative form to the frequently repeated but rarely formalized claim that AI ``strengthens every layer'' of a defense-in-depth architecture \cite{ejjami2024}.

\subsection{Human--AI Collaborative Triage Under a Capacity Constraint}\label{sec:triage}

The raw alert stream generated by Section~\ref{sec:cascade} arrives at rate $R = \pi_1 P_D(L) + (1 - \pi_1) P_F(L)$. An AI triage layer with threshold $\tAI$ prioritizes which alerts reach a human analyst of finite capacity $C$. Define the review probability $\prev = \min(1, C/R)$. Reviewed alerts pass through the analyst's own accuracy $(d_H, f_H)$; unreviewed alerts are auto-actioned at AI-only reliability:
\begin{equation}\label{eq:dfinal}
D_{\mathrm{final}} \;=\; P_D(L)\,\bigl[\prev\, d_H + (1 - \prev)\bigr]
\end{equation}
\begin{equation}\label{eq:ffinal}
F_{\mathrm{final}} \;=\; P_F(L)\,\bigl[\prev\, f_H + (1 - \prev)\bigr]
\end{equation}

Two structural properties follow directly. First, $F_{\mathrm{final}}$ strictly decreases as capacity increases (since $f_H < 1$), so human review's clearest value is filtering false alarms---reducing alert fatigue. Second, $D_{\mathrm{final}}$ can decrease as capacity increases whenever $d_H < 1$: an imperfect analyst reviewing every alert can be net-negative for detection relative to simply auto-actioning AI-flagged alerts, because the analyst's own error rate is then applied universally rather than to a filtered subset. This gives a precise, mechanistic reading of the reviewed literature's qualitative warning that over-reliance on any single layer---including the human layer---can leave a system worse off than a well-designed automated baseline \cite{ejeofobiri2024}.

Choosing $\tAI$ to maximize expected utility subject to the capacity constraint,
\begin{equation}\label{eq:utility}
\max_{\tAI}\; U(\tAI) \;=\; B\,\pi_1\, d(\tAI) \;-\; \cFP\,\pi_0\, f(\tAI) \;-\; c_H\, R(\tAI)
\quad \text{s.t.} \quad R(\tAI) \leq C
\end{equation}
gives, via the Lagrangian and the Neyman--Pearson structure of Section~\ref{sec:np}, an implicit equation for the optimal triage threshold $\tAI^{*}$ solvable by bisection on the Lagrange multiplier until the capacity constraint binds. We solve this numerically in Section~\ref{sec:sim}.

\subsection{Composite System Metric}\label{sec:composite}

Combining Sections~\ref{sec:cascade}--\ref{sec:triage} gives the full-pipeline detection probability:
\begin{equation}\label{eq:system}
D_{\mathrm{system}} \;=\; \left[\,1 - \prod_{i=1}^{L}\bigl(1 - d_i^{0}\bigr)\bigl(1 - \alpha_i\bigr)\right] \times \bigl[\prev\, d_H + (1 - \prev)\bigr]
\end{equation}
with a secondary Alert Fatigue Index, $\mathrm{AFI} = F_{\mathrm{final}} / C$, capturing operational sustainability---a quantity discussed qualitatively but not defined in the reviewed literature \cite{ejeofobiri2024,ejjami2024}.

\section{Simulation Methodology}\label{sec:sim}

The model of Section~\ref{sec:model} was implemented in Python (NumPy/SciPy) and evaluated through three experiments, using three illustrative Gaussian-separated layers representative of a network IDS, an endpoint EDR system, and an OT/SCADA anomaly monitor (parameters chosen for realistic but illustrative class separation; the simulation is a proof-of-concept rather than a calibration to a specific deployed system).

\paragraph{Experiment 1 --- Defense-in-Depth $\times$ AI augmentation.} $P_D(L)$ computed from Equations~\eqref{eq:pd}, \eqref{eq:aug}, and \eqref{eq:compound} for $L = 1, 2, 3$ layers, at AI augmentation levels $\alpha \in \{0, 0.3, 0.6\}$, with each layer's threshold set to a fixed 5\% false-positive rate via Equation~\eqref{eq:rates}.

\paragraph{Experiment 2 --- Pattern-recognition ROC.} Per-layer ROC curves and AUC computed by sweeping $\tau_i$ across each layer's Gaussian pair (Equation~\eqref{eq:rates}).

\paragraph{Experiment 3 --- Human--AI cascade.} For the fixed 3-layer system at $\alpha = 0.3$ ($\pi_1 = 0.02$ base rate, $d_H = 0.90$, $f_H = 0.05$), $D_{\mathrm{final}}$ and $F_{\mathrm{final}}$ (Equations~\eqref{eq:dfinal}--\eqref{eq:ffinal}) were computed across analyst capacity ratios $C/R \in [0.12, 1.00]$.

All code and exact parameter values are provided in the accompanying simulation script for reproducibility.

\section{Results}\label{sec:results}

\subsection{Defense-in-Depth $\times$ AI Augmentation}\label{sec:results-did}

\begin{figure}[ht]
\centering
\includegraphics[width=0.78\linewidth]{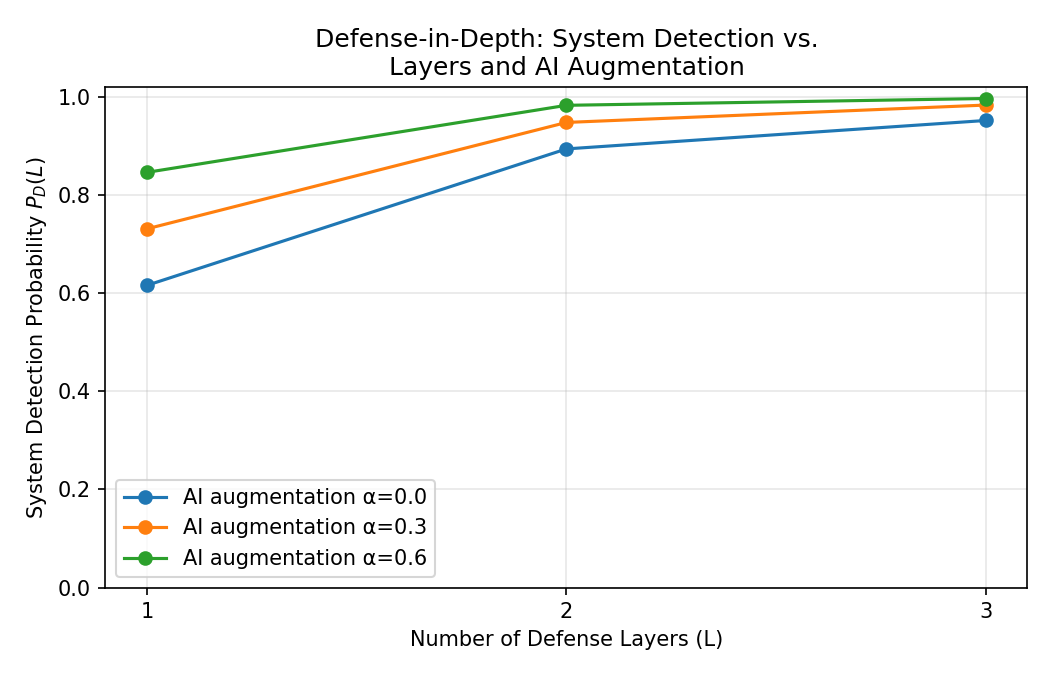}
\caption{System detection probability $P_D(L)$ versus number of defense layers, at three AI augmentation levels.}
\label{fig:did}
\end{figure}

Table~\ref{tab:did} reports the values plotted in Figure~\ref{fig:did}. Without AI ($\alpha = 0$), a single layer detects 61.6\% of attacks; three layers in series raise this to 95.2\%. AI augmentation compounds this effect: at $\alpha = 0.3$, three layers reach 98.4\%; at $\alpha = 0.6$, 99.7\%. Consistent with Equation~\eqref{eq:compound}, AI's marginal contribution is largest precisely where layering alone begins to saturate (the $L = 2 \to 3$ transition), rather than diminishing as layers alone do.

\begin{table}[ht]
\centering
\caption{System detection probability $P_D(L)$ by layer count and AI augmentation level.}
\label{tab:did}
\begin{tabular}{cccc}
\toprule
\textbf{Layers ($L$)} & \textbf{$\alpha = 0$ (no AI)} & \textbf{$\alpha = 0.3$} & \textbf{$\alpha = 0.6$} \\
\midrule
1 & 0.6164 & 0.7315 & 0.8465 \\
2 & 0.8942 & 0.9482 & 0.9831 \\
3 & 0.9521 & 0.9836 & 0.9969 \\
\bottomrule
\end{tabular}
\end{table}

\subsection{Pattern-Recognition ROC}\label{sec:results-roc}

\begin{figure}[ht]
\centering
\includegraphics[width=0.62\linewidth]{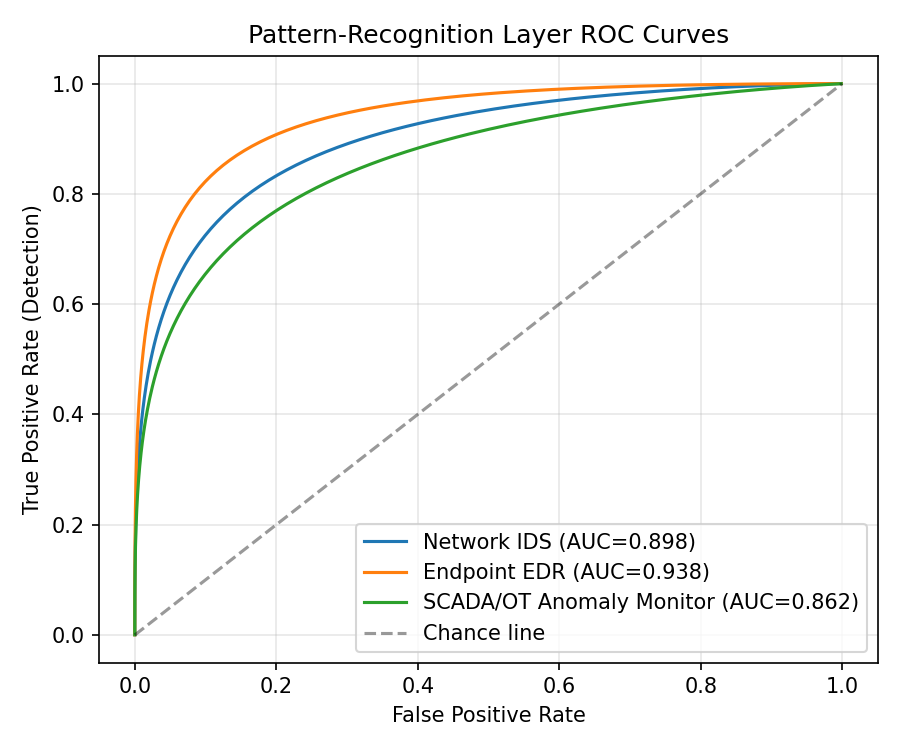}
\caption{ROC curves and AUC for each of the three illustrative defense layers.}
\label{fig:roc}
\end{figure}

Each layer's standalone ROC curve shows meaningful but incomplete discriminative power (AUC between 0.86 and 0.94 for the parameters used), underscoring why system-level detection probability $P_D(L)$---not any single detector---is the appropriate unit of analysis for defense-in-depth claims.

\subsection{Human--AI Collaborative Triage}\label{sec:results-triage}

\begin{figure}[ht]
\centering
\includegraphics[width=\linewidth]{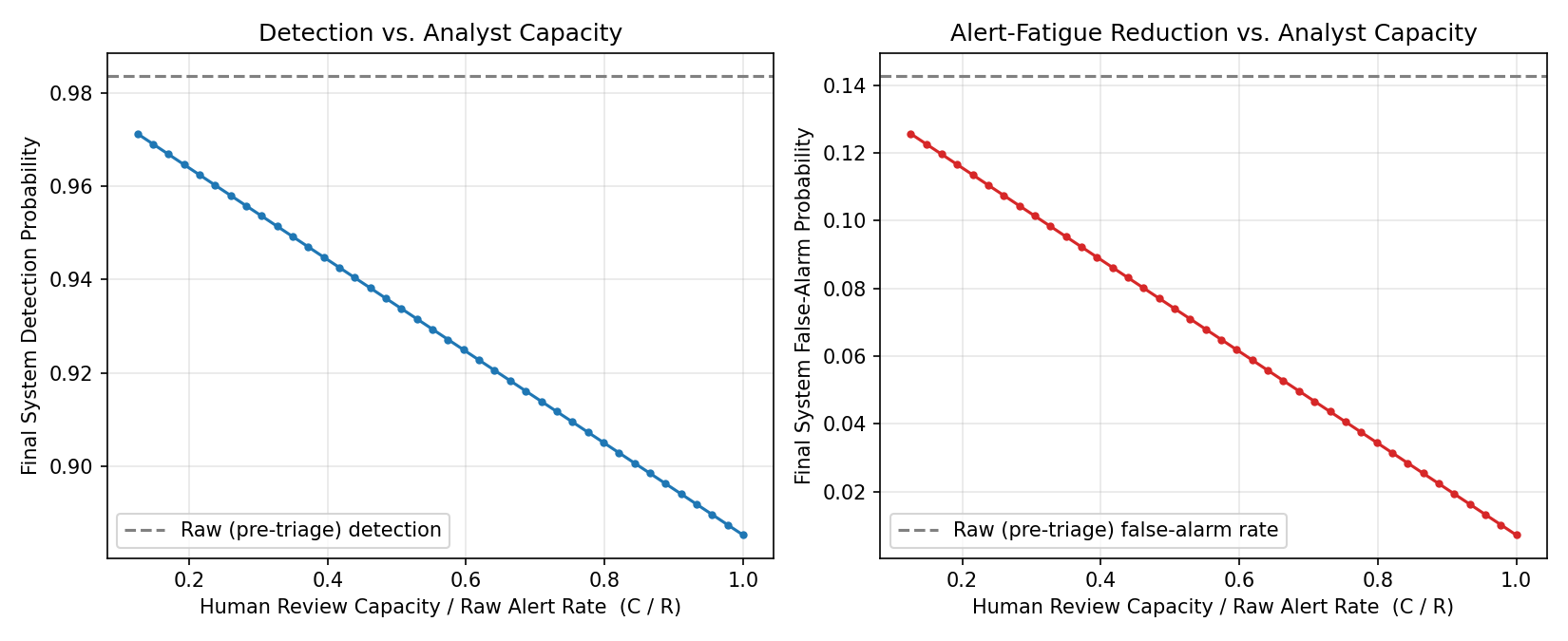}
\caption{Final system detection probability (left) and false-alarm probability (right) versus human analyst capacity ratio $C/R$, for the fixed 3-layer, $\alpha = 0.3$ system.}
\label{fig:triage}
\end{figure}

Table~\ref{tab:triage} summarizes the key operating points. Raw (pre-triage) system detection is 0.984 and raw false-alarm probability is 0.143. As analyst capacity increases from 12\% to 100\% of the raw alert volume, the false-alarm rate reaching action falls roughly 20-fold ($0.126 \to 0.007$), but detection probability falls from 0.971 to 0.885---a decline directly attributable to the analyst's own imperfect accuracy ($d_H = 0.90$) now being applied to every alert rather than a filtered subset.

\begin{table}[ht]
\centering
\caption{Final detection and false-alarm probabilities at selected analyst capacity ratios.}
\label{tab:triage}
\begin{tabular}{lcc}
\toprule
\textbf{$C/R$ (analyst capacity ratio)} & \textbf{Final detection $D_{\mathrm{final}}$} & \textbf{Final false-alarm $F_{\mathrm{final}}$} \\
\midrule
0.12 (raw, minimal review) & 0.971 & 0.126 \\
0.25 & 0.959 & 0.109 \\
1.00 (full review) & 0.885 & 0.007 \\
\bottomrule
\end{tabular}
\end{table}

\section{Discussion}\label{sec:discussion}

These results reframe two claims that recur across the reviewed literature in qualitative form, giving each a quantitative, testable structure.

\subsection{Defense-in-Depth Is Where AI Pays Off Most}\label{sec:disc-did}

Equation~\eqref{eq:compound} shows that AI's contribution to a layered architecture is multiplicative rather than additive: AI does not simply add a fixed increment of detection capability, it compounds the miss-probability reduction of every layer beneath it. Practically, this suggests that organizations gain more from deploying moderate AI augmentation ($\alpha \approx 0.3$) across several heterogeneous layers than from concentrating AI investment in a single, highly optimized detector---a specific, resource-allocation-relevant reading of the general claim that ``AI strengthens each layer of cybersecurity'' \cite{ejjami2024}.

\subsection{An Interior Optimum for Human--AI Collaboration}\label{sec:disc-interior}

The central and more novel finding is that full human review is not detection-optimal (Section~\ref{sec:results-triage}). This gives the frequently repeated recommendation to ``balance automation with human oversight'' \cite{ejeofobiri2024,ejjami2024} a specific, falsifiable content: the optimal analyst capacity ratio $C/R$ is an interior point determined by the analyst's own accuracy $d_H$ relative to 1, not an endpoint of ``more human review is always safer.'' Where $d_H$ is high (well-trained, unfatigued analysts), the model favors more review; where $d_H$ degrades---plausibly through the very automation-induced skill erosion the reviewed literature warns about \cite{ejeofobiri2024}---the model favors routing a larger share of alerts to direct AI action, creating a feedback loop worth investigating empirically in future SOC deployments.

\subsection{Implications for IT/OT-Converged Critical Infrastructure}\label{sec:disc-otit}

The three-layer configuration used here---network IDS, endpoint EDR, and an OT/SCADA anomaly monitor---mirrors the layer composition of security operations centers protecting IT/OT-converged environments such as oil and gas SCADA networks. In such environments, analyst capacity is often the binding constraint (specialized OT security expertise is scarce), making the capacity-constrained triage model of Section~\ref{sec:triage} directly actionable: Equation~\eqref{eq:utility} offers a principled way to set the AI triage threshold $\tAI$ given an organization's actual analyst headcount, rather than treating human review as an unlimited resource.

\section{Limitations and Future Work}\label{sec:limitations}

\paragraph{Synthetic, Gaussian-separated detectors.} The illustrative parameters are chosen for realistic but unverified class separation. Future work should calibrate $\mu$, $\sigma$, and $\alpha$ to empirical detector outputs on public IDS benchmark datasets (e.g., CICIDS2017, NSL-KDD) or to real SOC alert logs.

\paragraph{Layer independence.} Equations~\eqref{eq:miss}--\eqref{eq:pf} assume conditionally independent layers. Real attacks often evade correlated layers together (e.g., an attacker who defeats network-level detection may also be positioned to defeat endpoint detection). Extending the model with a copula or correlated-Bernoulli structure is a natural next step.

\paragraph{Static thresholds.} Thresholds $\tau_i$ and $\tAI$ are treated as fixed operating points; a reinforcement-learning or adaptive-control extension could let thresholds respond to observed attack-rate drift, addressing the adversarial-evasion concern raised in the reviewed literature \cite{ejeofobiri2024}.

\paragraph{Single triage layer and static analyst accuracy.} $d_H$ and $f_H$ are treated as constants; modeling their degradation under alert-volume-driven fatigue would let the model formally test the skill-erosion hypothesis discussed in Section~\ref{sec:disc-interior} rather than only motivating it.

\paragraph{Auto-action reliability.} Equations~\eqref{eq:dfinal}--\eqref{eq:ffinal} treat unreviewed AI-flagged alerts as fully actioned detections, which is what drives the decline of $D_{\mathrm{final}}$ with capacity. If automated response itself fails, is rate-limited, or is exploited by an adversary, the effective reliability of the unreviewed path falls below one and the interior optimum shifts back toward more human review; modeling that path explicitly is a direct extension.

\paragraph{Uniform review selection.} The review probability $\prev$ is applied uniformly to attack and benign alert streams, i.e., reviewed alerts are drawn as a random subset. Score-ordered triage---reviewing the highest-priority alerts first---would change the composition of the reviewed subset and generally improve on the uniform baseline reported here, making the present results a conservative bound on well-prioritized triage.

\paragraph{No empirical validation.} This paper establishes the model and demonstrates its qualitative and quantitative behavior through simulation; empirical validation against real SOC incident data is left for future work, ideally in collaboration with an operating SOC willing to share anonymized alert and triage-outcome data.

\section{Conclusion}\label{sec:conclusion}

This paper formalized two theoretical constructs widely used in the AI-cybersecurity review literature---Defense-in-Depth Theory and the Artificial Intelligence Theory of Pattern Recognition---as an explicit probabilistic cascade, and extended them with a capacity-constrained human--AI triage model. The resulting framework makes two claims precise that the literature otherwise states qualitatively: that AI augmentation compounds multiplicatively across defense layers, and that human oversight of AI-driven security systems has an interior optimum rather than a monotonic ``more is better'' relationship. Simulation results support both claims under illustrative but realistic parameters. The framework and its optimization structure (Equation~\eqref{eq:utility}) offer SOC designers---including those securing IT/OT-converged critical infrastructure---a concrete, falsifiable basis for allocating AI and human analyst resources, moving the field's guidance on responsible AI-augmented cyber defense from general principle toward testable design rule.

\bibliographystyle{unsrt}
\bibliography{references}

\end{document}